\documentclass[lettersize,journal]{IEEEtran}
\usepackage{array}
\usepackage[caption=false,font=normalsize,labelfont=sf,textfont=sf]{subfig}
\usepackage{stfloats}
\usepackage{url}
\usepackage{verbatim}
\usepackage{cite}
\usepackage{amsmath,amsfonts, amssymb, amsthm}  
\theoremstyle{remark}

\newtheorem{assumption}{Assumption}   
\newtheorem*{assumption*}{Assumption}   
\newtheorem{proposition}{Proposition}

\usepackage[T1]{fontenc}   

\usepackage{graphicx}
\usepackage{textcomp}
\usepackage{xcolor}

\usepackage{mathtools, mathcomp}  
\usepackage{comment}
\usepackage{mleftright}
\mleftright
\usepackage{bm}
\usepackage{mathtools}
\usepackage{xcolor}  
\usepackage{enumitem}  
\usepackage{enumerate}
\usepackage{flushend}  
\usepackage{multirow}
\usepackage{siunitx}

\usepackage{algorithmicx, algpseudocode, algorithm}  

\algrenewcommand\algorithmicindent{1.2em}  

\mathtoolsset{showonlyrefs=false}  

\usepackage{flushend}

\usepackage{lipsum}

\usepackage[]{hyperref}
\hypersetup{%
 setpagesize=false,%
 bookmarks=true,%
 bookmarksdepth=tocdepth,%
 bookmarksnumbered=true,%
 colorlinks=false,%
 pdftitle={},%
 pdfsubject={},%
 pdfauthor={},%
 pdfkeywords={}
 }

\begin{document}

\title{Sim-to-Real Powertrain Vibration Control via Model-Based Assisted Domain-Randomized Reinforcement Learning}

\author{Heisei Yonezawa,~\IEEEmembership{Member,~IEEE,}
        Ansei Yonezawa,~\IEEEmembership{Member,~IEEE,}
        Itsuro Kajiwara
\thanks{This work was supported in part by the Japan Society for the Promotion of Science (JSPS) KAKENHI [Grant Number 23K13273].}
\thanks{This work has been submitted to the IEEE for possible publication. Copyright may be transferred without notice, after which this version may no longer be accessible.}
}

\markboth{Journal of \LaTeX\ Class Files,~Vol.~xx, No.~xx, xxxx~20xx}%
{Shell \MakeLowercase{\textit{et al.}}: A Sample Article Using IEEEtran.cls for IEEE Journals}


\maketitle

\begin{abstract}
Controllers designed from nominal models often degrade on mechanical systems with nonlinearities and parametric variations, such as automotive powertrains with backlash. Deep reinforcement learning (DRL) with domain randomization can acquire robustness to such model discrepancies, but wide randomization destabilizes training and yields conservative policies, and experimental transfer of such policies has rarely been demonstrated in vibration control. This paper presents model-based controller assisted domain randomization (MBCADR), a framework that integrates model-based control into domain-randomized DRL. An $H_{2}$ controller designed from the nominal model supplies a baseline action, while a residual policy trained with twin delayed deep deterministic policy gradient under randomized dynamics compensates for nonlinearities and uncertainties. This model-based assistance relieves the agent of synthesizing the entire control input, thereby stabilizing training under extensive randomization. For each realization of the uncertain parameters, the augmented plant--controller state is shown to be a Markov state of the closed loop, so that domain-randomized training is rigorously formulated as policy optimization over a parameterized family of Markov decision processes. A Lyapunov-based stability analysis further derives a forward-invariant set and an explicit ultimate bound, in which the learned action enters only through its saturation level. Monte Carlo simulations with multiple training seeds show that MBCADR reduces performance variability compared with either component alone. Experiments on a powertrain test bench with backlash, using the simulation-trained policy on an embedded processor without further tuning, confirm successful sim-to-real transfer.
\end{abstract}


\section{Introduction}   \label{S_Introduction}

\subsection{Background and Motivation}   \label{S_Background_and_Motivation}
\IEEEPARstart{I}{ndustrial} mechanical systems, such as automotive powertrains \cite{Wang2023,Wang2022} and robotic manipulators \cite{Wu2024,Zaare2022}, are subject to increasingly stringent performance requirements, and their dynamics have become correspondingly complex. Nonlinearities, time delays, and variations in physical parameters are inherent in such systems \cite{Yang2025,Xu2022}, and controllers must ensure stability and performance in their presence. A representative example is the automotive powertrain, in which the torque generated by the actuator is transmitted to the wheels through compliant shafts and gears, so that abrupt torque transients excite torsional vibration that degrades drivability. Gear backlash further aggravates this problem by inducing switching between contact and non-contact modes, which renders the dynamics discontinuous \cite{Yonezawa2022,Yonezawa2024a}.

Since the advent of state-space theory \cite{Kalman1960}, model-based control has been the mainstay of industrial practice, and its performance is fundamentally governed by model fidelity. Robust control methods such as $H_{\infty}$ control \cite{Zhou1996,Xu2024,Chen2026, Li2026} guarantee performance under uncertainty, but only for uncertainties that can be represented in the form required by the synthesis. Discontinuous nonlinearities such as backlash, and simultaneous variations in multiple parameters, are difficult to capture in this form without conservative designs. Moreover, modeling and calibration errors can never be fully eliminated, so a controller that performs well in simulation may degrade when deployed on the physical plant. This problem is known as the sim-to-real gap \cite{Peng2018}.

Deep reinforcement learning (DRL) offers an alternative that does not require an explicit transition model, since the agent improves its policy from data collected through interaction with the environment \cite{Sutton2018,Mnih2013,Mnih2015}. DRL has been applied to vibration suppression of flexible link manipulators \cite{Viswanadhapalli2024}, flexible rotors \cite{Ahmed2023}, magnetorheological elastomer absorbers \cite{Wang2024}, flexible hinged plates \cite{Qiu2024}, and robot control \cite{Malmir2026}. It has also been combined with conventional controllers, for example with a proportional–integral–derivative (PID) controller \cite{Rehan2026} and with active disturbance rejection control \cite{Chen2023}.

Training a DRL agent directly on hardware is rarely practical: the exploratory actions required for learning may drive the plant into unsafe operating regions, and the required volume of interaction data is costly to acquire \cite{Peng2018,Slawik2024}. Agents are therefore commonly trained in simulation, where exploration is safe and data are virtually unlimited. A policy trained in this manner, however, is just as susceptible to the sim-to-real gap as a model-based controller.

\subsection{Related Work}   \label{S_Related_Works}
Domain randomization (DR) addresses the sim-to-real gap by randomizing the physical parameters of the simulator at the start of each training episode \cite{Peng2018,Zhang2023}. The policy is then optimized over a set of Markov decision processes (MDPs) rather than a single one, which promotes robustness to parameter variations within the randomized range. A theoretical analysis has also clarified the conditions under which DR achieves successful transfer \cite{Chen2022}. DR has been applied successfully to object pushing with a 7-DOF manipulator \cite{Peng2018}, deformable object manipulation \cite{Matas2018}, bipedal locomotion \cite{Li2021,Castillo2026}, legged locomotion learned from animal motion \cite{Peng2020}, humanoid robots \cite{Liao2024,Gu2024}, unmanned aerial vehicle \cite{Yu2026}, path following of underwater vehicles \cite{Fan2024}, and overhead crane control under payload variations \cite{Zhang2023}.

Previous DR studies leave two issues from a control-engineering standpoint. First, DR has been applied predominantly in robotics, whereas its use for vibration control of mechanical systems remains scarce. Systems with discontinuous nonlinearities, such as powertrains with backlash, have been addressed only in a few studies, including the authors' previous work \cite{Yonezawa2025}. Second, because the training environment changes from episode to episode, wide randomization makes the learning problem harder \cite{Matas2018}, destabilizes training, and tends to produce overly conservative policies \cite{Nachum2019,Ramos2019,Cheng2022}. Enlarging the network and the amount of training data can offset these effects in principle, but at the cost of longer development, a higher risk of overfitting, and policies that are too large for the memory-constrained embedded processors on which automotive control laws are executed.

Combining a model-based controller with a learning component is a promising way to address these issues. This idea was formalized as residual reinforcement learning by Johannink \textit{et al.} \cite{Johannink2019} and Silver \textit{et al.} \cite{Silver2018}, in which a policy learns a corrective action that is added to the output of an existing controller. Johannink \textit{et al.} showed on a real robot that this structure accomplishes a block assembly task involving contacts that are difficult to model. Silver \textit{et al.} showed that residual learning substantially improves imperfect controllers in complex manipulation tasks for which learning from scratch is data-inefficient. Koryakovskiy \textit{et al.} \cite{Koryakovskiy2018} used reinforcement learning to compensate for the model--plant mismatch of a nonlinear model predictive controller and validated the approach on a real robot. Other combinations of model-based control and reinforcement learning have also been reported \cite{Okawa2019, Park2023}. Cao \textit{et al.} \cite{Cao2023} combined linear state feedback with a residual policy trained using a Lyapunov-like reward, and derived conditions under which a safety envelope is rendered invariant and the closed loop is stabilized. Although some of these studies introduce variations in task conditions, a formulation as DR that substantially varies the dynamics of the controlled plant itself has not been investigated in depth. Moreover, the application of these hybrid schemes to automotive systems and vibration control remains largely unexplored.

Liu \textit{et al.} \cite{Liu2022} combined a PID controller with a residual policy for autonomous blimp control, randomized wind and buoyancy during training, and transferred the simulation-trained policy to a real blimp. Zuo \textit{et al.} \cite{Zuo2023} replaced the PID controller with an $H_{\infty}$ controller to enlarge the stability margin, and derived a lower bound on the mixing factor between the two actions that ensures stability under assumptions on the distribution of the learned actions. Kim \textit{et al.} \cite{Kim2026} combined nonlinear model predictive control and a disturbance observer with a residual policy trained under randomized payloads and friction, and bounded the learned action on the basis of an input-to-state stability analysis. Yan \textit{et al.} \cite{Yan2026} augmented a model-based bipedal locomotion controller with a residual policy trained under dynamics randomization, and accelerated learning through a supervised loss derived from a model-based oracle with privileged access to the true dynamics. However, these hybrid schemes are formulated for a particular plant and a particular baseline controller, often at the cost of a more complex control structure with additional components such as disturbance observers, estimator networks, and privileged information; a general formulation of the domain-randomized optimization problem for the resulting closed loop has not been established. Moreover, the stability of the closed loop including the learned policy has been addressed only in limited settings.

In the authors' previous work, Yonezawa \textit{et al.} \cite{Yonezawa2025} proposed a hybrid approach combining domain-randomized DRL and output-feedback $H_{2}$ vibration control and showed in simulation that it achieves robustness with a compact network and limited training data. A preliminary version of this work \cite{Yonezawa2026CCTA} introduced TD3 into the framework. Both studies, however, were confined to numerical simulation, and two questions remained open. First, it was not shown whether adding a learned action to a model-based closed loop preserves the stability that the model-based controller provides. Second, it was not shown whether the robustness obtained through DR carries over to physical hardware when the policy runs on an embedded processor.

\subsection{Contributions of This Study}   \label{S_Contributions}
\begin{figure}[htbp]
    \centering
    \includegraphics[width=0.5\textwidth]{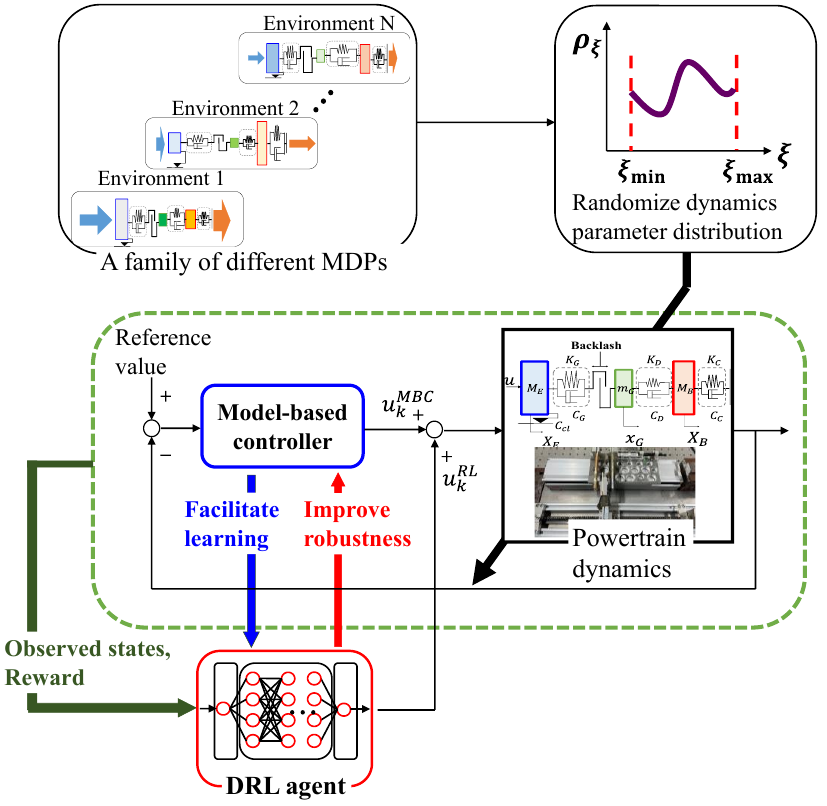}
    \caption{Outline of the proposed approach.}
    \label{Fig_Overview}
\end{figure}

This paper proposes model-based controller assisted domain randomization (MBCADR), a robust control framework that integrates model-based control into domain-randomized DRL, as outlined in Fig.~\ref{Fig_Overview}. DRL does not replace the model-based controller. Instead, it learns corrective actions for modeling errors, while the model-based controller provides the baseline action from the start of training. With this division of roles, the agent only needs to learn compensation for model discrepancies rather than the entire control input. This reduction of the learning burden becomes increasingly important as the randomization range widens, because wider randomization makes training harder. Although an automotive powertrain is used as the application example, the framework is formulated for general nonlinear systems with state-space model-based controllers. This paper extends the preliminary version \cite{Yonezawa2026CCTA} with a closed-loop stability analysis, Monte Carlo evaluation over multiple training seeds, and experimental validation on a physical test bench.

The main contributions are as follows:
\begin{itemize}
\item[1)] \textit{Framework and formulation:} A simple control framework that strengthens DR through the assistance of a model-based controller designed from a nominal model. For each realization of the uncertain parameters, the augmented plant--controller state is shown to satisfy the Markov property. Domain-randomized training is therefore formulated as policy optimization over a family of MDPs parameterized by the uncertain physical parameters.
\item[2)] \textit{Stability analysis:} A Lyapunov-based analysis of the hybrid closed loop derives a forward-invariant set and an explicit ultimate bound. These are computed in advance from the design setting, without modifying the learned policy online. The learned action enters the bound only through its saturation level, so the trade-off between learning authority and the guaranteed bound can be evaluated in advance for any internally stabilizing model-based controller.
\item[3)] \textit{Experimental sim-to-real validation:} Monte Carlo simulations with multiple training seeds show that MBCADR reduces performance variability compared with model-based control alone and with DR alone. Experiments on a powertrain test bench with backlash under several hardware configurations confirm successful sim-to-real transfer. The simulation-trained policy, a compact network that fits in the memory of an embedded processor, runs on the processor without further tuning.
\end{itemize}

The remainder of this paper is organized as follows. Section~\ref{S_Problem_Formulation} formulates the control problem. Section~\ref{S_RL_Preliminaries} reviews the MDP and TD3. Section~\ref{S_Proposed_Approach} presents MBCADR. Section~\ref{S_Stability_Analysis} analyzes the stability of the hybrid closed loop. Sections~\ref{S_Numerical_Verifications} and \ref{S_Experiment} report the simulation and experimental results, respectively. Section~\ref{S_Conclusion} concludes the paper.

\section{Problem Formulation}   \label{S_Problem_Formulation}
Consider a discrete-time nonlinear system whose dynamics depend on uncertain physical parameters:
\begin{equation}   \label{Eq:x+=f(x,u,w;xi)}
    x_{k+1} = f \left(x_{k}, u_{k}, w_{k}; \xi \right)
\end{equation}
\begin{equation}   \label{Eq:y=h(x;xi)}
    y_{k} = h \left(x_{k}; \xi \right)
\end{equation}
where $x_{k} \in \mathbb{R}^{n}$, $u_{k} \in \mathbb{R}^{m}$, $w_{k} \in \mathbb{R}^{l}$, and $y_{k} \in \mathbb{R}^{o}$ denote the state, control input, exogenous disturbance, and controlled output at time step $k$, respectively. The maps $f \colon \mathbb{R}^{n} \times \mathbb{R}^{m} \times \mathbb{R}^{l} \to \mathbb{R}^{n}$ and $h \colon \mathbb{R}^{n} \to \mathbb{R}^{o}$ describe the state transition and the output, and neither is known exactly to the designer. The vector $\xi$ collects the uncertain physical parameters of the plant, such as inertial and damping parameters and the width of a dead zone, which also govern the nonlinear effects. In the simulator used for training, the dynamics are parameterized by the same vector $\xi$, so that each training episode is specified by a realization of $\xi$ (see Section~\ref{S_DR_and_Augumented_Dynamics}).

The control objective is output tracking: for a given reference $y_{k}^{r} \in \mathbb{R}^{o}$, the tracking error $e_{k} := y_{k}^{r} - y_{k}$ should converge to zero, that is, $\lim_{k \to \infty} \| e_{k} \| = 0$. This objective is cast as the finite-horizon optimal control problem
\begin{equation}   \label{Eq:OCP}
    \begin{aligned}
        \underset{\left\{u_{0},u_{1},\ldots,u_{T}\right\} \in U}{\min} &\quad \sum_{k = 0}^{T} c \left(x_{k}, u_{k} \right) \\
        \mathrm{subject \; to} &\quad 
        \eqref{Eq:x+=f(x,u,w;xi)}, \eqref{Eq:y=h(x;xi)}, \ k = 0,1,\ldots,T
    \end{aligned}
\end{equation}
where $U$ is the admissible input set. The stage cost $c \colon \mathbb{R}^{n} \times \mathbb{R}^{m} \to \mathbb{R}$ is nonnegative, vanishes when $e_{k} = 0$ and $u_{k} = 0$, and grows unboundedly as $\| (e_{k}, u_{k}) \| \to \infty$. In this study, the quadratic cost
\begin{equation}   \label{Eq:StageCost}
    c \left(x_{k}, u_{k} \right) = e_{k}^{\top} Q e_{k} + u_{k}^{\top} R u_{k}
\end{equation}
is adopted, where $Q \in \mathbb{R}^{o \times o}$ is positive semidefinite and $R \in \mathbb{R}^{m \times m}$ is positive definite.

Because $f$ and $h$ are not known exactly, \eqref{Eq:OCP} cannot be solved directly by conventional model-based design. In many mechanical systems, however, partial knowledge of the dynamics is available. The following assumptions formalize this knowledge.

\begin{assumption}[Nominal model]   \label{Assum:Nominal_Model}
A linear nominal model of \eqref{Eq:x+=f(x,u,w;xi)} and \eqref{Eq:y=h(x;xi)} is available:
\begin{equation}   \label{Eq:x+=Ax+B1w+B2u}
    x_{k+1} = Ax_{k} + B_{1} w_{k} + B_{2} u_{k}
\end{equation}
\begin{equation}   \label{Eq:y=Cx}
    y_{k} = Cx_{k}
\end{equation}
where $A \in \mathbb{R}^{n \times n}$, $B_{1} \in \mathbb{R}^{n \times l}$, $B_{2} \in \mathbb{R}^{n \times m}$, and $C \in \mathbb{R}^{o \times n}$. Such a model can be obtained from first-principles modeling or system identification at nominal parameter values.
\end{assumption}

\begin{assumption}[Model-based controller]   \label{Assum:MBC}
A discrete-time linear controller of the form
\begin{equation}   \label{Eq:StateEquation_MBC}
    x_{k+1}^{c} = A_{c}x_{k}^{c} + B_{c}e_{k}
\end{equation}
\begin{equation}   \label{Eq:OutputEquation_MBC}
    u_{k}^{MBC} = C_{c}x_{k}^{c} + D_{c}e_{k}
\end{equation}
with internal state $x_{k}^{c} \in \mathbb{R}^{n_{c}}$ and matrices $A_{c} \in \mathbb{R}^{n_{c} \times n_{c}}$, $B_{c} \in \mathbb{R}^{n_{c} \times o}$, $C_{c} \in \mathbb{R}^{m \times n_{c}}$, and $D_{c} \in \mathbb{R}^{m \times o}$, is available such that the feedback interconnection of \eqref{Eq:x+=Ax+B1w+B2u}--\eqref{Eq:OutputEquation_MBC} is internally stable.
\end{assumption}

Assumption~\ref{Assum:MBC} specifies only the structure of the controller; any synthesis method that yields an internally stabilizing controller in the form of \eqref{Eq:StateEquation_MBC} and \eqref{Eq:OutputEquation_MBC} may be used. The specific controller adopted in this study is described in Section~\ref{S_Hybrid_Control_Input}.

The discrepancy between the true dynamics \eqref{Eq:x+=f(x,u,w;xi)} and the nominal model \eqref{Eq:x+=Ax+B1w+B2u} is denoted by
\begin{equation}   \label{Eq:Model_Mismatch}
    \Delta_{k} := f \left(x_{k}, u_{k}, w_{k}; \xi \right) - \left(Ax_{k} + B_{1} w_{k} + B_{2} u_{k} \right).
\end{equation}
This term collects the effects neglected in \eqref{Eq:x+=Ax+B1w+B2u}, namely nonlinearities such as backlash and the deviations of $\xi$ from the nominal values. Because the controller \eqref{Eq:StateEquation_MBC} and \eqref{Eq:OutputEquation_MBC} is designed without knowledge of $\Delta_{k}$, its tracking performance can degrade, and stability can even be lost, when it is applied to the true plant \eqref{Eq:x+=f(x,u,w;xi)} and \eqref{Eq:y=h(x;xi)}. The objective of this paper is to retain the model-based controller \eqref{Eq:StateEquation_MBC} and \eqref{Eq:OutputEquation_MBC} while compensating for $\Delta_{k}$ through learning.

\section{Reinforcement Learning Preliminaries}   \label{S_RL_Preliminaries}

\subsection{Markov Decision Process}   \label{S_MDP}
Reinforcement learning problems are formulated as a Markov decision process (MDP) defined by the tuple $\left(\mathcal{S}, \mathcal{A}, p, r, \gamma \right)$, where $\mathcal{S}$ denotes the state space, $\mathcal{A}$ the action space, $p \left(s_{k+1} \mid s_{k},a_{k} \right)$ the state transition probability, $r\left(s_{k},a_{k}\right)$ the reward function, and $0 < \gamma < 1$ the discount factor \cite{Sutton2018,Silver2014,Lillicrap2019}. At each discrete time step $k$, the agent observes a state $s_{k}$, selects an action $a_{k}$ according to a policy $\pi$, and receives a reward $r_{k}$. The objective is to maximize the expected cumulative discounted reward
\begin{equation}   \label{Eq:CumReward}
    J\left(\pi\right) = \mathbb{E} \left[ \sum_{i=1}^{T} \gamma^{i-1} r\left(s_{i},a_{i}\right) \mid s_1 = s,a_1 = a; \pi\right]
\end{equation}
In this study, the reward is defined from the stage cost \eqref{Eq:StageCost} with the opposite sign, so that maximizing \eqref{Eq:CumReward} corresponds to minimizing the cost in \eqref{Eq:OCP}.

\subsection{Twin Delayed Deep Deterministic Policy Gradient}   \label{S_Twin_Delayed_DDPG}
For continuous-action control problems, a deterministic policy can be learned by actor--critic methods using off-policy data \cite{Silver2014}, where a critic estimates the value function. In this study, Twin Delayed Deep Deterministic Policy Gradient (TD3) \cite{Fujimoto2018} is employed as the learning algorithm for the reinforcement learning component. TD3 uses a deterministic policy (\emph{actor}) $\mu_{\theta}\left(\cdot\right)$ and two action-value functions (\emph{critics}) $Q_{\phi_{1}}\left(\cdot\right)$ and $Q_{\phi_{2}}\left(\cdot\right)$. The actor outputs a continuous action $a_{k} = \mu_{\theta}\left(s_{k} \right)$. To achieve sample-efficient learning, TD3 is trained off-policy using transitions stored in a replay buffer $\mathcal{D}$. To reduce overestimation bias, TD3 maintains two critics and uses a conservative target based on their minimum. Given a sampled transition $\left(s_{k},a_{k},r_{k},s_{k+1} \right) \sim \mathcal{D}$, the target action is computed using a slowly-updated target actor $\mu_{\theta -}$ with a smoothed perturbation:
\begin{equation}   \label{Eq:Update_TargetActor}
    \begin{gathered}   
        \tilde{a}_{k+1} = \mu_{\theta -} \left(s_{k+1} \right) + \varepsilon \\
        \varepsilon \sim \mathrm{clip}\left(\mathcal{N}\left(0,\sigma^{2}\right),-c,c\right).
    \end{gathered}
\end{equation}
Then, the target value is defined as
\begin{equation}   \label{Eq:TargetValue}
    Y_{k} = r_{k} + \gamma \underset{i\in\left\{1,2\right\}}{\min} \ Q_{\phi_{i}^{-}} \left(s_{k+1}, \tilde{a}_{k+1} \right)
\end{equation}
where $Q_{\phi_{i}^{-}}$ denotes target critics. Each critic is updated by minimizing the temporal-difference loss
\begin{multline}
    L \left(\phi_{i}\right) 
    = \mathbb{E}_{\left(s_{k},a_{k},r_{k},s_{k+1} \right) \sim \mathcal{D}} \left[\left(Q_{\phi_{i}} \left(s_{k}, a_{k} \right) - Y_{k}\right)^{2} \right], \\
      i \in \left\{1,2\right\}.
\end{multline}
The actor is updated less frequently than the critics, i.e., delayed policy updates, to stabilize the critic output. Using the chain rule, the deterministic policy gradient is computed as
\begin{equation}   \label{Eq:DDPG}
    \nabla _{\theta} J \left( \mu_{\theta} \right) 
    = \mathbb{E}_{s_{k} \sim \mathcal{D}} \left[\nabla _{\theta} \mu_{\theta} \left( s_{k} \right) \nabla_{a}Q_{\phi_{min}} \mid _{a=\mu_{\theta} \left( s_{k} \right)} \right]
\end{equation}
where $Q_{\phi_{min}}$ denotes the critic that attains $\min\left(Q_{\phi_{1}}, Q_{\phi_{2}}\right)$ at $\left(s_{k}, \mu_{\theta} \left( s_{k} \right)\right)$.

\section{Proposed Approach}   \label{S_Proposed_Approach}

\subsection{Model-Based Controller and Hybrid Control Input}   \label{S_Hybrid_Control_Input}
In this study, the controller in Assumption~\ref{Assum:MBC} is an $H_{2}$ controller synthesized from the nominal model by the LMI approach \cite{Chilali1996,Zhou1996}. Any other controller satisfying Assumption~\ref{Assum:MBC} can be used without changing the subsequent formulation. To compensate for the model mismatch $\Delta_{k}$, a learning-based input is added to the model-based input, and the overall control input is defined as
\begin{equation}   \label{Eq:Hybrid_Control_Input}
    u_{k} = u_{k}^{MBC} + u_{k}^{RL}
\end{equation}
where $u_{k}^{RL}$ is generated by a reinforcement learning policy.

The two components in \eqref{Eq:Hybrid_Control_Input} play complementary roles. From the viewpoint of the learning agent, the MBC supplies a stabilizing action derived from the nominal model from the first training episode. Because the closed loop already achieves reasonable tracking before learning starts, the agent avoids a large number of uninformative trials that an immature policy would otherwise generate, and exploration is concentrated around a meaningful operating region. This assistance is particularly valuable under DR, in which the environment changes from episode to episode and learning the entire control law from scratch becomes difficult \cite{Matas2018}. From the viewpoint of the MBC, the learning agent compensates for the model mismatch $\Delta_{k}$ that the MBC cannot address because it is designed only from the nominal model.

This division of roles reduces the burden on the learning agent: the agent focuses on compensating for residual dynamics rather than synthesizing the entire control input, so that the function to be approximated by the policy becomes simpler. As a result, a compact network and a limited amount of training data suffice \cite{Yonezawa2025}. Moreover, because the nominal performance is secured by the MBC, the policy is not required to achieve robustness at the expense of the baseline performance, which is expected to mitigate the conservativeness that DR tends to induce \cite{Nachum2019,Ramos2019,Cheng2022}. Finally, the learned correction $u_{k}^{RL}$ is bounded by the saturating output layer of the actor, so that it acts only within a prescribed range around the baseline action; this property is exploited in the stability analysis of Section~\ref{S_Stability_Analysis}.

\subsection{Domain Randomization and Augmented Dynamics}   \label{S_DR_and_Augumented_Dynamics}
During training, domain randomization is employed to address parametric uncertainty. At the beginning of each episode, the parameter vector $\xi$ is randomly sampled from a predefined distribution $\rho_{\xi}$ and remains fixed throughout the episode \cite{Peng2018,Zhang2023}.

Standard reinforcement learning algorithms, including TD3, are derived under the premise that the environment is an MDP, that is, the next state depends only on the current state and action \cite{Sutton2018}. It is therefore desirable that the training environment satisfies this property \cite{Okawa2019}. Under the hybrid input \eqref{Eq:Hybrid_Control_Input}, however, the agent does not interact with the plant alone: the MBC, which has its own internal state, is inserted between the agent and the plant. It must therefore be confirmed that the environment seen by the agent remains an MDP.

From the viewpoint of the agent, the nonlinear plant and the MBC together constitute a single environment, as illustrated in Fig.~\ref{Fig_Overview}. The agent supplies $u_{k}^{RL}$ to this closed-loop system and receives signals generated within it. Consequently, not only the plant output but also internal signals of the MBC can be included in the observation of the agent. The state of this environment is obtained by augmenting the plant state with the controller state:
\begin{equation}   \label{Eq:State_Environment}
    \mathcal{X}_{k}^{env} := \begin{bmatrix}x_{k}^{\top} & x_{k}^{c\top}   \end{bmatrix} ^{\top},
\end{equation}
and the closed-loop dynamics under the hybrid control input \cite{Yonezawa2025} are written as
\begin{equation}    \label{Eq:Closed-loop}
    \begin{aligned}
        & \mathcal{X}_{k+1}^{env} = \begin{bmatrix}x_{k+1} \\ x_{k+1}^{c} \end{bmatrix} 
        = \begin{bmatrix}f \left(x_{k},u_{k},w_{k};\xi\right) \\ A_{c}x_{k}^{c} + B_{c}e_{k} \end{bmatrix} \\
        & = \begin{bmatrix}f \left(x_{k}, u_{k}^{RL} + u_{k}^{MBC},w_{k};\xi\right) \\ A_{c}x_{k}^{c} + B_{c}e_{k} \end{bmatrix} \\
        & = \begin{bmatrix}f \left(x_{k},u_{k}^{RL} + C_{c}x_{k}^{c} + D_{c} \left(y_{k}^{r} - y_{k}\right),w_{k};\xi\right) \\ A_{c}x_{k}^{c} + B_{c}\left(y_{k}^{r} - y_{k}\right) \end{bmatrix} \\
        & = \begin{bmatrix}f \left(x_{k},u_{k}^{RL} + C_{c}x_{k}^{c} + D_{c} y_{k}^{r} - D_{c}h\left(x_{k};\xi\right),w_{k};\xi\right) \\ A_{c}x_{k}^{c} + B_{c}y_{k}^{r} - B_{c}h\left(x_{k};\xi\right) \end{bmatrix} \\
        & := \mathcal{F}^{env}\left(\mathcal{X}_{k}^{env}, u_{k}^{RL}, w_{k}, y_{k}^{r}; \xi\right).
    \end{aligned}
\end{equation}
The parameter vector $\xi$ is fixed within each episode, and $w_{k}$ and $y_{k}^{r}$ are exogenous signals that are independent of the past states and inputs. Hence, \eqref{Eq:Closed-loop} shows that, for a fixed realization of $\xi$, the distribution of the next augmented state $\mathcal{X}_{k+1}^{env}$ depends only on the current augmented state $\mathcal{X}_{k}^{env}$ and the learning-based input $u_{k}^{RL}$. The closed loop therefore satisfies the Markov property for each parameter realization, and constitutes an MDP whose state is $\mathcal{X}_{k}^{env}$ and whose action is $u_{k}^{RL}$ \cite{Okawa2019,Koryakovskiy2018}.

This result has two practical consequences. First, the RL algorithms and DR techniques developed for MDPs can be applied to the hybrid scheme without modification, because the MBC is simply absorbed into the environment. Second, the derivation uses only the state-space structure of the MBC in Assumption~\ref{Assum:MBC} and the general form of the plant \eqref{Eq:x+=f(x,u,w;xi)} and \eqref{Eq:y=h(x;xi)}; the result therefore holds for any such plant and controller, not only for the powertrain and the $H_{2}$ controller considered in this study. The family of these closed-loop MDPs indexed by $\xi$ forms the basis of the policy optimization in Section~\ref{S_Latent_MDP_Formulation}.

\subsection{Policy Optimization over a Family of MDPs}   \label{S_Latent_MDP_Formulation}
Under domain randomization, the training environment is no longer a single MDP. As shown in Section~\ref{S_DR_and_Augumented_Dynamics}, each realization of $\xi$ yields a closed-loop MDP, denoted by $\mathcal{M}\left( \xi \right)$, whose state is $\mathcal{X}_{k}^{env}$ and whose action is $u_{k}^{RL}$. The simulator can therefore be regarded as a family of MDPs $\mathcal{M} = \left\{ \mathcal{M}\left( \xi \right) \right\}$ parameterized by the uncertain physical parameters, together with the distribution $\rho_{\xi}$ from which the member used in each episode is randomly sampled \cite{Zhang2023,Chen2022}.

Let $\mu_{\theta}$ be a policy parameterized by $\theta$, and let $\tau_{1:T}$ denote a trajectory generated in $\mathcal{M}\left( \xi \right)$. The expected return under domain randomization is defined as
\begin{equation}   \label{Eq:DR_Loss}
    \begin{multlined}
        J\left(\mu_{\theta} \right) \\
        = \mathbb{E}_{\xi \sim \rho_{\xi}} \left[ 
            \mathbb{E}_{\tau \sim p \left(\tau \mid \mu_{\theta},\xi\right)} \left[
                \sum_{i=1}^{T} \gamma^{i-1} r \left(s_{i}, a_{i} \right) \mid s_{1},a_{1}; \mu_{\theta}
            \right]
         \right]
    \end{multlined}
\end{equation}
where the likelihood of a trajectory $\tau = \left(s_1, a_1, s_2, \ldots, a_{T-1}, s_T\right)$ is denoted by $p \left(\tau | \mu_{\theta},\xi \right)$ under the policy $\mu_{\theta}$.
The learning objective is given by
\begin{equation}   \label{Eq:DR_Objective}
    \mu_{\theta}^{\ast} = \underset{\mu_{\theta}}{\mathrm{arg \, max}} \; J\left( \mu_{\theta} \right).
\end{equation}

The objective \eqref{Eq:DR_Loss} evaluates a policy by its average return over the family rather than by its return in a single member. The distribution $\rho_{\xi}$ thus specifies the range of variations against which robustness is sought: a wider $\rho_{\xi}$ covers larger model discrepancies, at the cost of a harder learning problem and potentially more conservative policies, as discussed in Section~\ref{S_Introduction}. The model-based assistance described in Section~\ref{S_Hybrid_Control_Input} is introduced to alleviate this trade-off. The overall procedure is summarized in Algorithm~\ref{Algorithm_TD3-MBCA-DRT}.

\begin{algorithm}[!b]
  \caption{Model-Based Controller Assisted Domain-Randomized Training}
      \label{Algorithm_TD3-MBCA-DRT}
  \begin{algorithmic}[1]

    \Require{Environment $E$ and trajectories pool $\mathcal{T}_{rand}$.}   
    \Ensure{Optimal actor policy.}  
    
    \State Randomly initialize the actor's policy parameters and the critic parameters.
    \State Design a MBC using the nominal system model.
    
    \For{each episode}
      \State $\xi \sim \rho_{\xi}$: Randomly sample a set of physical parameters.
      \State $E \gets \Xi\left(\xi\right)$: Reset the simulation environment
      \State Generate rollout $\tau = \left(s_{1},a_{1}, \ldots, s_{T}\right) \sim \mu_{\theta} + u_{MBC}$ 
      \Statex \hspace{0.8em} with dynamics $\xi$ and $E$.
      \State $\mathcal{T}_{rand} \gets \mathcal{T}_{rand} \cup \tau_{i}$
      \State $u_{RL} \gets \mu_{\theta}$: DRL control action from the policy.
      \State $u = u_{RL} + u_{MBC}$: Integration of the data-driven policy 
      \Statex \hspace{0.8em} $u_{RL}$ and the physical model-based control $u_{MBC}$.
        \For{each gradient step}
          \State Update the critic parameters by minimizing the
          \Statex \hspace{2.1em} mean squared error.
          \State Update the actor parameters using the sampled
          \Statex \hspace{2.1em} policy gradient.
          \State $\theta \gets \theta + \alpha_{\theta} \nabla_{\theta} J\left(\mu_{\theta} \right)$ with $\mathcal{T}_{rand}$ update.
          \State $u = u_{RL} + u_{MBC}$: Update the hybrid control. 
        \EndFor
    \EndFor
  \end{algorithmic}
\end{algorithm}

\section{Stability Analysis}   \label{S_Stability_Analysis}
This section analyzes the stability of the hybrid closed loop in continuous time, in which the MBC is designed. The analysis starts from the generalized plant description, so that the result applies to any internally stabilizing model-based controller rather than to a particular synthesis route.


Let the nominal plant used for the MBC design be written in generalized plant form:
\begin{equation}
\begin{aligned}
\dot{x} &= A_{p}x + B_{p1}w + B_{p2}u \\
z &= C_{p1}x + D_{p12}u \\
y &= C_{p2}x + D_{p21}w
\end{aligned}
\label{eq:genplant}
\end{equation}
where $w$ collects the reference signal $y^{r}$ and exogenous disturbances, $z$ is the performance output, and $y$ is the signal fed back to the controller; the tracking error $e = y^{r} - C_{p}x$ is recovered by an appropriate choice of $C_{p2}$ and $D_{p21}$. The standard assumptions $D_{p11} = 0$ and $D_{p22} = 0$ are adopted. The MBC (\ref{Eq:StateEquation_MBC}) and (\ref{Eq:OutputEquation_MBC}), written in continuous time, is
\begin{equation}
\dot{x}_{c} = A_{K}x_{c} + B_{K}y, \qquad
u^{MBC} = C_{K}x_{c} + D_{K}y
\label{eq:mbc_ct}
\end{equation} 
The true plant departs from the nominal model because of the backlash nonlinearity and the parametric variations. This mismatch is treated as an additive unmodeled term $\Delta(t)$ acting on the state equation, rather than being embedded in the generalized plant as a linear fractional term:
\begin{equation}
\dot{x} = A_{p}x + B_{p1}w + B_{p2}u + \Delta
\label{eq:trueplant}
\end{equation}
Substituting $y = C_{p2}x + D_{p21}w$ into (\ref{eq:mbc_ct}) and applying the hybrid input (\ref{Eq:Hybrid_Control_Input}) gives $u = C_{K}x_{c} + D_{K}(C_{p2}x + D_{p21}w) + u^{RL}$. With the augmented state $\mathcal{X} := [\, x^{\top} \;\; x_{c}^{\top} \,]^{\top}$, the closed loop becomes
\begin{equation}
\dot{\mathcal{X}} = A_{cl}\mathcal{X} + B_{cl}w
+ \tilde{B}_{2}u^{RL} + \tilde{B}_{\Delta}\Delta
\label{eq:closedloop}
\end{equation}
where
\begin{equation}
\begin{aligned}
A_{cl} &= \begin{bmatrix}
A_{p} + B_{p2}D_{K}C_{p2} & B_{p2}C_{K} \\
B_{K}C_{p2} & A_{K}
\end{bmatrix}, \\[2pt]
B_{cl} &= \begin{bmatrix}
B_{p1} + B_{p2}D_{K}D_{p21} \\
B_{K}D_{p21}
\end{bmatrix}, \\[2pt]
\tilde{B}_{2} &= \begin{bmatrix} B_{p2} \\ 0 \end{bmatrix}, \quad
\tilde{B}_{\Delta} = \begin{bmatrix} I \\ 0 \end{bmatrix}.
\end{aligned}
\label{eq:clmatrices}
\end{equation}

\begin{assumption}[Bounded signals]
\label{Assum:Bounded}
The exogenous input, the learned action, and the model mismatch are bounded as $\|w(t)\| \le \bar{w}$, $\|u^{RL}(t)\| \le \bar{u}$, and $\|\Delta(t)\| \le \bar{\Delta}$.
\end{assumption}

The bound on $u^{RL}$ holds by construction, since the actor terminates in a saturating output layer; the bound on $\Delta$ holds on the operating region because the dead-zone force is limited by the drivetrain stiffness and the backlash width, and the parametric variations lie in the ranges of domain randomization.

\begin{proposition}
\label{prop:ub}
Suppose that Assumption~\ref{Assum:Bounded} holds. Since the MBC internally stabilizes the nominal plant, $A_{cl}$ is stable, and for any
$S = S^{\top} \succ 0$ the Lyapunov equation
\begin{equation}
A_{cl}^{\top}P + PA_{cl} = -S
\label{eq:lyap}
\end{equation}
has a unique solution $P = P^{\top} \succ 0$. Let
\begin{equation}
\begin{aligned}
b &:= \|PB_{cl}\|\bar{w} + \|P\tilde{B}_{2}\|\bar{u}
+ \|P\tilde{B}_{\Delta}\|\bar{\Delta}, \\
\bar{\rho} &:= \frac{2b}{\lambda_{\min}(S)}
\end{aligned}
\label{eq:bdef}
\end{equation}
and define the sublevel set of
$V(\mathcal{X}) = \mathcal{X}^{\top}P\mathcal{X}$
\begin{equation}
\Omega := \left\{ \mathcal{X} :
V(\mathcal{X}) \le \lambda_{\max}(P)\,\bar{\rho}^{\,2} \right\}.
\label{eq:omega}
\end{equation}
Then $\Omega$ is forward invariant, every trajectory asymptotically approaches $\Omega$, and the closed loop state is ultimately bounded as
\begin{equation}
\limsup_{t \to \infty} \|\mathcal{X}(t)\|
\le \sqrt{\kappa(P)}\;\bar{\rho}, \qquad
\kappa(P) := \frac{\lambda_{\max}(P)}{\lambda_{\min}(P)}.
\label{eq:ultimatebound}
\end{equation}
\end{proposition}

\begin{IEEEproof} 
Let $V(\mathcal{X}) = \mathcal{X}^{\top}P\mathcal{X}$. Differentiating along (\ref{eq:closedloop}) and using $P = P^{\top}$ to merge the two cross terms, then substituting (\ref{eq:lyap}), gives
\begin{align}
\dot{V} &= \mathcal{X}^{\top}\!\left(A_{cl}^{\top}P + PA_{cl}\right)\!
\mathcal{X}
+ 2\mathcal{X}^{\top}P\!\left(B_{cl}w + \tilde{B}_{2}u^{RL}
+ \tilde{B}_{\Delta}\Delta\right) \notag \\
&= -\mathcal{X}^{\top}S\mathcal{X}
+ 2\mathcal{X}^{\top}P\!\left(B_{cl}w + \tilde{B}_{2}u^{RL}
+ \tilde{B}_{\Delta}\Delta\right)
\label{eq:vdot}
\end{align}
Applying the Rayleigh quotient to the first term, and the Cauchy--Schwarz inequality to each contribution in the second, yields
\begin{align}
\dot{V} &\le -\lambda_{\min}(S)\|\mathcal{X}\|^{2}
+ 2b\|\mathcal{X}\| \notag \\
&= -\lambda_{\min}(S)\,\|\mathcal{X}\|
\left(\|\mathcal{X}\| - \frac{2b}{\lambda_{\min}(S)}\right)
\label{eq:vdotbound}
\end{align}
Hence $\dot{V} < 0$ holds whenever $\|\mathcal{X}\| > \bar{\rho} := 2b/\lambda_{\min}(S)$. 

The set $\Omega$ in (\ref{eq:omega}) is the smallest sublevel set of $V$ containing the ball $\mathcal{B} := \{\|\mathcal{X}\| \le \bar{\rho}\}$: for $\mathcal{X} \in \mathcal{B}$, the Rayleigh quotient gives $V(\mathcal{X}) \le \lambda_{\max}(P)\|\mathcal{X}\|^{2} \le \lambda_{\max}(P)\bar{\rho}^{\,2}$, and the bound is attained on $\mathcal{B}$. Consequently $\mathcal{X} \notin \Omega$ implies $\mathcal{X} \notin \mathcal{B}$, that is, $\|\mathcal{X}\| > \bar{\rho}$, and therefore $\dot{V} < 0$ outside $\Omega$. On the boundary of $\Omega$ one has $\lambda_{\max}(P)\|\mathcal{X}\|^{2} \ge V(\mathcal{X}) = \lambda_{\max}(P)\bar{\rho}^{\,2}$, hence $\|\mathcal{X}\| \ge \bar{\rho}$ and $\dot{V} \le 0$; the trajectory cannot cross the boundary outward, so $\Omega$ is forward invariant \cite{khalil2002}.

\textcolor{black}
{
Consequently $V$ decreases until the trajectory enters the smallest sublevel set of $V$ containing the ball $\{\|\mathcal{X}\| \le \bar{\rho}\}$, and cannot subsequently leave it, since $\dot{V} \le 0$ on its boundary. The bound (\ref{eq:ultimatebound}) then follows from
$\lambda_{\min}(P)\|\mathcal{X}\|^{2} \le V(\mathcal{X})
\le \lambda_{\max}(P)\|\mathcal{X}\|^{2}$ \cite{khalil2002}.
}
\end{IEEEproof}

Proposition 1 characterizes the region to which the hybrid closed loop is ultimately confined, and quantifies the contribution of the learning-based input to the Lyapunov decrease condition of the closed loop. The learning-based input enters $\bar{\rho}$ only through the single term $\|P\tilde{B}_{2}\|\bar{u}$, which is linear in the saturation bound $\bar{u}$. The learned policy therefore cannot destabilize the closed loop; its effect is limited to an enlargement of the ultimate bound (\ref{eq:ultimatebound}) that is computable in advance from $\bar{u}$ and the design data. As $\bar{u} \to 0$ the expression reduces to the corresponding evaluation for the closed loop driven by $w$ and $\Delta$ alone, so $\bar{u}$ acts as an explicit design parameter trading learning authority against the size of the guaranteed ultimate bound. The three contributions to $\bar{\rho}$ are separated by physical origin---the reference and disturbance channel, the learning-based input, and the model mismatch---and can be evaluated individually for a given design.

\section{Numerical Verifications}   \label{S_Numerical_Verifications}

\subsection{Nonlinear Powertrain Model}   \label{S_Powertrain}
\begin{figure}[htbp]
    \centering
    \includegraphics[width=0.48\textwidth]{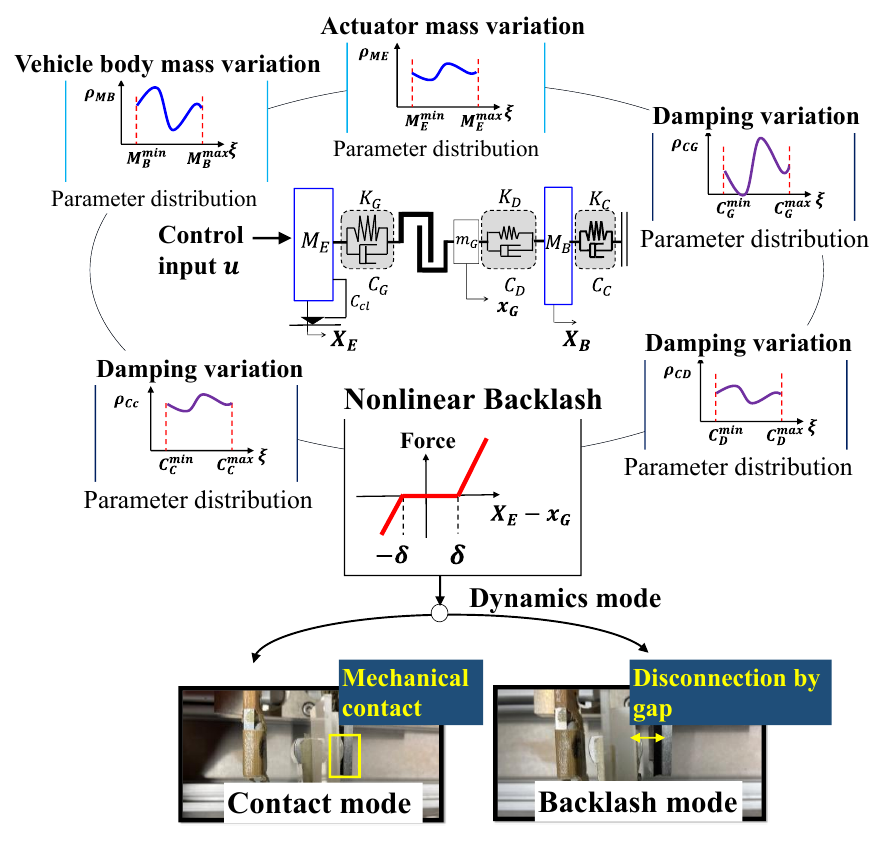}
    \caption{Powertrain dynamics.} 
    \label{Fig_Powertrain}
\end{figure}

Numerical simulations are conducted on a nonlinear vehicle powertrain system with mechanical backlash and parametric uncertainty. The backlash introduces a dead-zone nonlinearity, resulting in mode-dependent dynamics that complicate vibration suppression. A schematic overview of the equivalent simple powertrain model is shown in Fig. \ref{Fig_Powertrain}. \textcolor{black}{See the reference \cite{Yonezawa2025}\cite{Yonezawa2022} for more details on the powertrain model and the nominal parameters.}
The controlled output is defined as the vehicle body vibration $y_{k} = X_{B}$, and the control objective is to suppress vibration while tracking a reference signal under nonlinear and uncertain operating conditions.

\subsection{Verification Settings}   \label{S_Verification_Setting}
The verification settings consist of learning conditions and uncertainty conditions, which are summarized separately.

The hyperparameters and implementation settings used for training the learning-based controller are summarized in Table \ref{Table_TD3_Parameter}. These settings are fixed throughout all simulations to ensure consistent evaluation. The proposed method is compared with three baselines. The first is the open-loop system without feedback control (No control). The second is the model-based controller alone (Only MBC), that is, the $H_{2}$ controller described in Section~\ref{S_Hybrid_Control_Input} without the learning-based input. The third is domain-randomized DRL without model-based assistance (Only DR), in which the policy generates the entire control input. The proposed method and Only DR are trained under the same conditions listed in Table~\ref{Table_TD3_Parameter}, except for the components related to the model-based controller. Each learning-based method is trained with five random seeds. For application to powertrain control, a feedforward input is also involved \cite{Yonezawa2019a,Yonezawa2024a}.

\begin{figure}[htbp]
    \centering
    \includegraphics[width=0.45\textwidth]{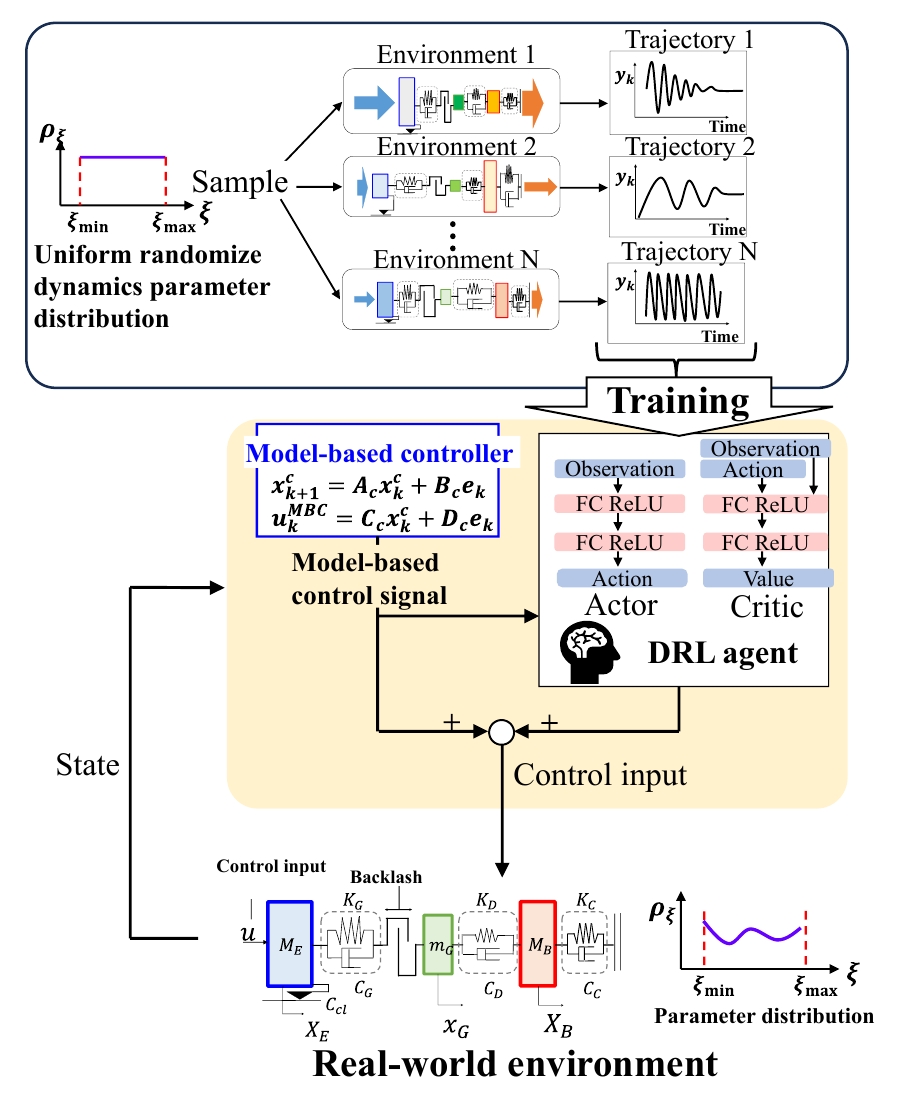}
    \caption{Outline of domain randomized training.} 
    \label{Fig_Training}
\end{figure}

Parametric uncertainty is introduced through domain randomization. At the beginning of each episode, uncertain physical parameters are randomly sampled and kept fixed during the episode. The ranges of the randomized parameters, including the actuator mass $M_{E} \in \left[M_{E}^{min}, M_{E}^{max} \right]$, vehicle body mass $M_{B} \in \left[M_{B}^{min}, M_{B}^{max} \right]$, three damping coefficients $C_{G} \in \left[C_{G}^{min}, C_{G}^{max}\right]$, $C_{D} \in \left[C_{D}^{min}, C_{D}^{max}\right]$, $C_{C} \in \left[C_{C}^{min}, C_{C}^{max}\right]$, and backlash length $\delta \in \left[\delta^{min}, \delta^{max} \right]$, are summarized in Table \ref{Table_Randomized_Parameters}. Figure \ref{Fig_Training} outlines the training procedure. 

\begin{table}[htbp]
\caption{Hyperparameters for TD3.}
\begin{center}
    {\renewcommand\arraystretch{1.2}  
        \tabcolsep = 5.0pt   
\begin{tabular}{|c|c|}
\cline{1-2}
 \emph{\textbf{Hyperparameter}} & \emph{\textbf{Value}}  \\ \cline{1-2}
 Sampling time & $\qty{0.006}{\second}$  \\ \cline{1-2}
 Critic learning rate & $1.0 \times 10^{-4}$  \\ \cline{1-2}
 Actor learning rate & $1.0 \times 10^{-4}$  \\ \cline{1-2}
 Discount factor $\gamma$ & $0.99$ \\ \cline{1-2}
 \begin{tabular}{c}Number of the neurons in \\ the hidden layers \end{tabular} & $64$ \\ \cline{1-2}
 Size of the mini-batch $M$ & $128$ \\ \cline{1-2}
 Size of the replay buffer & $1.0 \times 10^{5}$ \\ \cline{1-2}
 Horizon $T$ & $667$ \\ \cline{1-2}
 Network architecture & Fully connected + ReLU \\ \cline{1-2}
 Number of episodes & $500$ \\ \cline{1-2}
 Target smoothing noise & $\mathrm{clip} \left(\mathcal{N}\left(0,0.2\right),-0.5,0.5\right)$  \\ \cline{1-2}
 Exploration noise & Ornstein--Uhlenbeck action noise  \\ \cline{1-2}
 \begin{tabular}{c}Frequency of policy update \\ with respect to critic update \end{tabular} & $2$ \\ \cline{1-2}
\end{tabular}
    }
\label{Table_TD3_Parameter}
\end{center}
\end{table}

\begin{table}[htbp]
\caption{Randomized parameters.}
\begin{center}
    {\renewcommand\arraystretch{1.2}  
        \tabcolsep = 3.0pt   
\begin{tabular}{|c|l|}
\cline{1-2}
 \emph{\textbf{Parameter}} & \emph{\textbf{Range}}  \\ \cline{1-2}
 Damping coefficient $C_D$ & \begin{tabular}{l}Nominal value $\pm 50 \%$: \\ $\left[C_{D}^{min}, C_{D}^{max} \right] = \left[6.25,18.75 \right]$ \end{tabular} \\ \cline{1-2}
 Damping coefficient $C_G$ & \begin{tabular}{l}Nominal value $\pm 50 \%$: \\ $\left[C_{G}^{min}, C_{G}^{max} \right] = \left[18,54 \right]$ \end{tabular} \\ \cline{1-2}
 Damping coefficient $C_C$ & \begin{tabular}{l}Nominal value $\pm 50 \%$: \\ $\left[C_{C}^{min}, C_{C}^{max} \right] = \left[0.05,0.15 \right]$ \end{tabular} \\ \cline{1-2}
 Vehicle body mass $M_B$ & \begin{tabular}{l}Nominal value $\pm 50 \%$: \\ $\left[M_{B}^{min}, M_{B}^{max} \right] = \left[0.1160,0.3480 \right]$ \end{tabular} \\ \cline{1-2}
 Actuator mass $M_E$ & \begin{tabular}{l}Nominal value $\pm 50 \%$: \\ $\left[M_{E}^{min}, M_{E}^{max} \right] = \left[0.5200,1.5600 \right]$ \end{tabular} \\ \cline{1-2}
 Length of backlash $\delta$ & \begin{tabular}{l}Nominal value $\pm 50 \%$: \\ $\left[\delta^{min}, \delta^{max} \right] = \left[0.0025,0.0075 \right]$ \end{tabular}\\ \cline{1-2}
 \begin{tabular}{c}Steady value of the reference \\ signal from $0$ to $\qty{2}{s}$ $y_{0-2s}^{r}$ \end{tabular} & \begin{tabular}{l}$\left[y_{0-2s}^{r,min}, y_{0-2s}^{r,max} \right]$ \\ $ = \left[-0.01515,0.030303\right] \, \si{m}$ \end{tabular}  \\ \cline{1-2}
 \begin{tabular}{c}Steady value of the reference \\ signal from $2$ to $\qty{4}{s}$ $y_{2-4s}^{r}$ \end{tabular} & \begin{tabular}{l}$\left[y_{2-4s}^{r,min}, y_{2-4s}^{r,max} \right]$ \\ $ = \left[-0.01515,0.030303\right] \, \si{m}$ \end{tabular}  \\ \cline{1-2}
\end{tabular}
    }
\label{Table_Randomized_Parameters}
\end{center}
\end{table}

The observation vector includes the reference signal, the controlled output, the tracking error and its integral and derivative, as well as the control input generated by the model-based controller. All observation signals are normalized. The reward function is defined as a quadratic cost penalizing tracking error and control effort, consistent with the problem formulation in Section \ref{S_Problem_Formulation}.

\subsection{Results and Discussion}   \label{S_Results_and_Discussion}
\begin{figure}[htbp]
    \centering
    \includegraphics[width=0.35\textwidth]{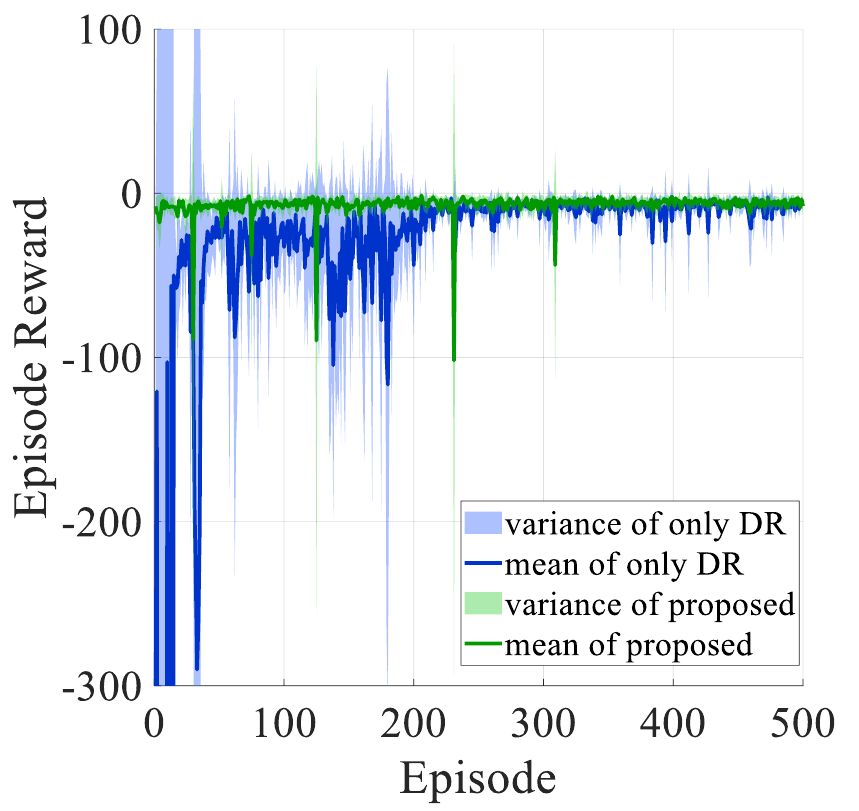}
    \caption{Episode reward during training over five seeds.}
    \label{Fig_Reward}
\end{figure}

Fig.~\ref{Fig_Reward} shows the episode reward during training over the five seeds. For Only DR (blue line), the learning curves vary considerably across seeds. In contrast, the proposed method (green line) yields learning curves with markedly smaller variation, suggesting that introducing the MBC stabilizes training. Because the early stage of training is dominated by trial-and-error exploration, a DRL agent that generates the entire control input is prone to destabilizing the closed loop, and the convergence of Only DR is therefore delayed compared with that of the proposed method. By running the MBC in parallel, the agent of the proposed method can avoid unproductive exploration, which improves convergence.

Table~\ref{Table_Result_Nominal} compares the control performance for the nominal model, which contains neither nonlinearities nor parametric uncertainty, using the $2$-norm of the error between the reference and the controlled output. For the learning-based methods (the proposed method and Only DR), the mean and standard deviation over the five seeds are reported. All three closed-loop systems reduce the error substantially compared with No control, and Only MBC achieves the best performance. Since the model used for the controller design and the plant coincide, it is theoretically reasonable that the MBC performs best. In the proposed method, this nominal performance of the MBC serves as the foundation that assists the training of the DRL agent.

\begin{table}[htbp]
\caption{$2$-norm of the tracking error for the nominal model (mean and standard deviation over five seeds)}
\begin{center}
    {\renewcommand\arraystretch{1.2}
        \tabcolsep = 8.0pt
\begin{tabular}{|c|c|c|}
\hline
 \emph{\textbf{Method}} & \emph{\textbf{Mean}} & \emph{\textbf{Standard deviation}} \\ \hline
 Proposed method & $0.7960$ & $0.1458$ \\ \hline
 Only DR & $0.9788$ & $0.1787$ \\ \hline
 Only MBC & $0.5890$ & -- \\ \hline
 No control & $2.6940$ & -- \\ \hline
\end{tabular}
    }
\label{Table_Result_Nominal}
\end{center}
\end{table}

\begin{figure}[htbp]
    \centering
    \includegraphics[width=\columnwidth]{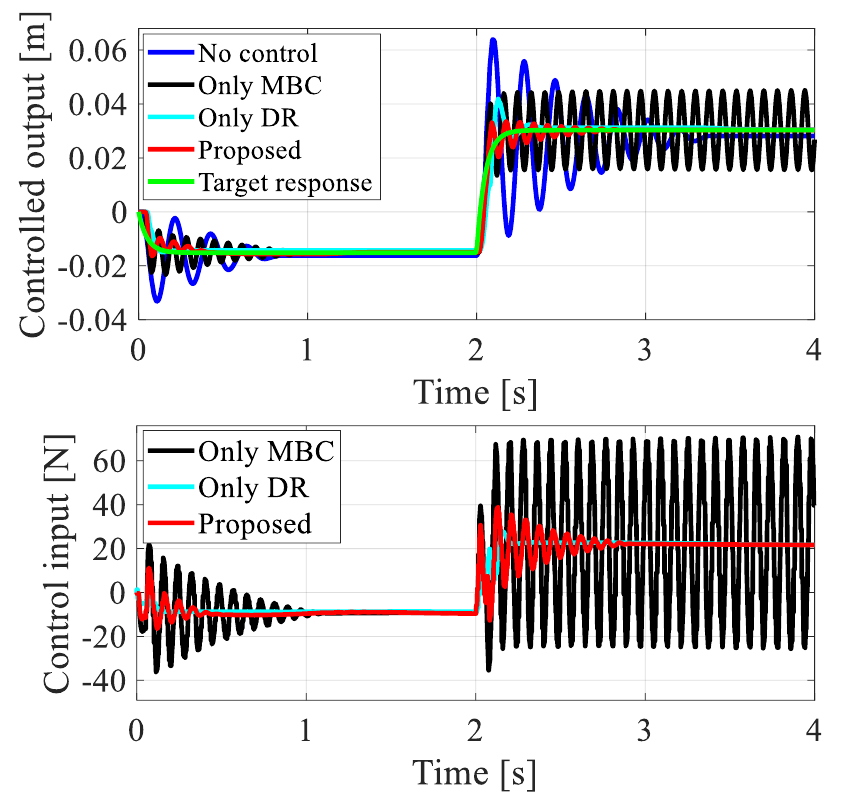}
    \caption{Simulation result with $M_{B} = M_{B}^{min}$, $M_{E} = M_{E}^{min}$, $y_{0-2s}^{r} = y_{0-2s}^{r,min}$, $y_{2-4s}^{r} = y_{2-4s}^{r,max}$, $C_{G} = C_{G}^{max}$, $C_{D} = C_{D}^{min}$, $C_{C} = C_{C}^{min}$, and $\delta = \delta^{max}$.}
    \label{Fig_Robustness_Mmin}
\end{figure}

\begin{figure}[htbp]
    \centering
    \includegraphics[width=\columnwidth]{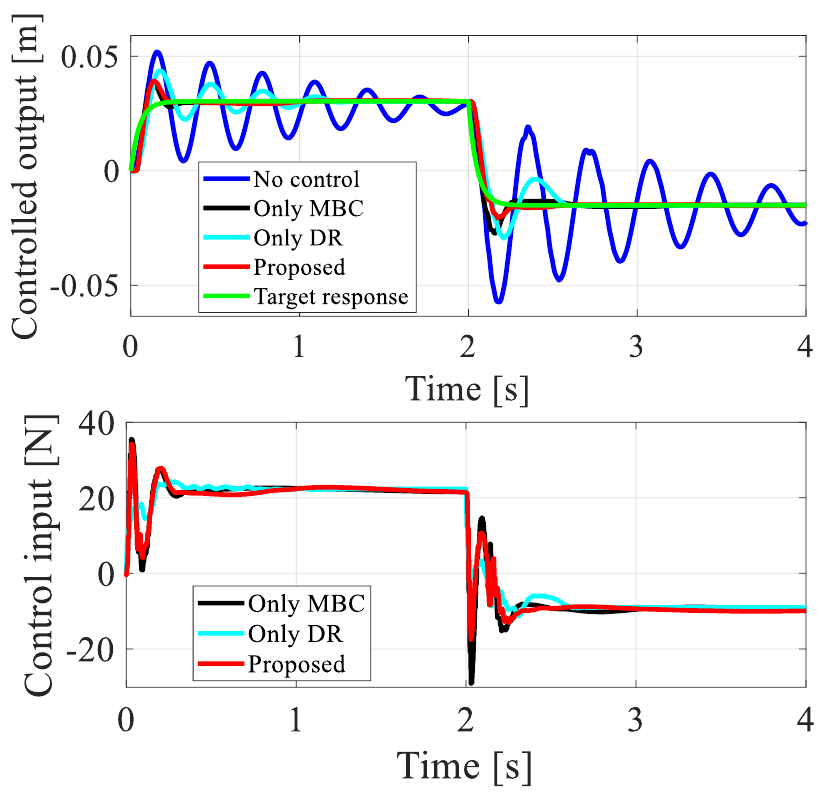}
    \caption{Simulation result with $M_{B} = M_{B}^{max}$, $M_{E} = M_{E}^{max}$, $y_{0-2s}^{r} = y_{0-2s}^{r,max}$, $y_{2-4s}^{r} = y_{2-4s}^{r,min}$, $C_{G} = C_{G}^{min}$, $C_{D} = C_{D}^{max}$, $C_{C} = C_{C}^{max}$, and $\delta = \delta^{min}$.}
    \label{Fig_Robustness_Max}
\end{figure}

Next, robustness against plant uncertainty is examined. Figs.~\ref{Fig_Robustness_Mmin} and~\ref{Fig_Robustness_Max} show representative results when the powertrain dynamics include parameter variations and the backlash nonlinearity. In these figures, multiple parameters, such as the vehicle mass, are set to the extremes of their ranges, with opposite combinations in the two cases. The upper and lower panels show the time histories of the vehicle body vibration and the control input, respectively. The ideal response (green line) contains abrupt changes in the driving force, under which vibration is most likely to be excited. Tables~\ref{Table_Result_Robustness_Mmin} and~\ref{Table_Result_Robustness_Max} compare the $2$-norms for the results in Figs.~\ref{Fig_Robustness_Mmin} and~\ref{Fig_Robustness_Max}, respectively, where the mean and standard deviation are reported for the learning-based methods.

\begin{table}[htbp]
\caption{$2$-norm of the tracking error for Fig.~\ref{Fig_Robustness_Mmin} (mean and standard deviation over five seeds)}
\begin{center}
    {\renewcommand\arraystretch{1.2}
        \tabcolsep = 8.0pt
\begin{tabular}{|c|c|c|}
\hline
 \emph{\textbf{Method}} & \emph{\textbf{Mean}} & \emph{\textbf{Standard deviation}} \\ \hline
 Proposed method & $1.1863$ & $0.1477$ \\ \hline
 Only DR & $1.3640$ & $0.4619$ \\ \hline
 Only MBC & $3.3755$ & -- \\ \hline
 No control & $4.1192$ & -- \\ \hline
\end{tabular}
    }
\label{Table_Result_Robustness_Mmin}
\end{center}
\end{table}

\begin{table}[htbp]
\caption{$2$-norm of the tracking error for Fig.~\ref{Fig_Robustness_Max} (mean and standard deviation over five seeds)}
\begin{center}
    {\renewcommand\arraystretch{1.2}
        \tabcolsep = 8.0pt
\begin{tabular}{|c|c|c|}
\hline
 \emph{\textbf{Method}} & \emph{\textbf{Mean}} & \emph{\textbf{Standard deviation}} \\ \hline
 Proposed method & $1.6472$ & $0.1802$ \\ \hline
 Only DR & $2.6016$ & $1.0243$ \\ \hline
 Only MBC & $1.4891$ & -- \\ \hline
 No control & $6.3969$ & -- \\ \hline
\end{tabular}
    }
\label{Table_Result_Robustness_Max}
\end{center}
\end{table}

A notable finding of these robustness tests is that the proposed method (MBC + DRL with DR) reduces the variability with respect to both the training conditions and the plant variations. In Tables~\ref{Table_Result_Robustness_Mmin} and~\ref{Table_Result_Robustness_Max}, the spread of the norm across seeds is substantially smaller for the proposed method than for Only DR. This indicates that the MBC can ensure the consistency of the policy against the stochastic uncertainty inherent in DRL with DR. DR is a major source of the diversity of the generated trajectories; the MBC is therefore particularly beneficial in training in which the exploration space tends to expand. The weak vibration suppression of Only DR observed in Fig.~\ref{Fig_Robustness_Max} is attributed to a suboptimal policy produced by excessive DR. The residual vibration that remains even with the proposed method in Fig.~\ref{Fig_Robustness_Mmin} suggests that the degree of DR in this study is too large. Nevertheless, the comparison of the mean $2$-norms in Tables~\ref{Table_Result_Robustness_Mmin} and~\ref{Table_Result_Robustness_Max} shows that the proposed method alleviates this conservativeness to some extent by introducing the MBC, because the MBC secures a baseline performance from the beginning of training. In other words, the proposed method is considered to enable learning focused on the uncertainty by efficiently restricting the space to be explored.

The comparison between the proposed method and Only MBC also indicates the robustness of the proposed method. Only MBC sometimes outperforms the proposed method, as in Table~\ref{Table_Result_Robustness_Max}, but its performance can also deteriorate severely, as in Table~\ref{Table_Result_Robustness_Mmin}. Because the MBC is vulnerable to modeling errors, it cannot provide consistent robustness over diverse plants; the persistent vibration in Fig.~\ref{Fig_Robustness_Mmin} is a typical example. Although the proposed method shows somewhat insufficient performance in Fig.~\ref{Fig_Robustness_Mmin} and Table~\ref{Table_Result_Robustness_Max}, it can be regarded as broadly robust against various variations, owing to the improved generalization provided by DR.

To evaluate the robustness statistically, two further evaluations are conducted. The first is a Monte Carlo evaluation over plants obtained from 100 random samples. Since 100 samples are generated for each of the five training seeds, 500 closed-loop simulations are performed for each learning-based method. The mean and standard deviation of the $2$-norm of the tracking error over these runs are summarized in Table~\ref{Table_Result_MonteCarlo}.

The second is an extreme-parameter evaluation, which examines the performance at the boundary of the randomized ranges. Each method is evaluated at all $2^{8} = 256$ vertices of the ranges listed in Table~\ref{Table_Randomized_Parameters}, that is, combinations of the minimum and maximum values of the six physical parameters and the two reference levels. For the learning-based methods, this evaluation is repeated for each of the five trained policies, resulting in 1280 simulations per method. The mean and standard deviation of the $2$-norm over these runs are summarized in Table~\ref{Table_Result_ExtremeParameter}. Whereas the Monte Carlo evaluation characterizes the typical performance over the training distribution, the extreme-parameter evaluation examines the performance under extreme parameter combinations at the boundary of this distribution, where controllers are most likely to degrade.

\begin{table}[htbp]
\caption{$2$-norm of the tracking error in the Monte Carlo evaluation (100 trials for each of five seeds)}
\begin{center}
    {\renewcommand\arraystretch{1.2}
        \tabcolsep = 8.0pt
\begin{tabular}{|c|c|c|}
\hline
 \emph{\textbf{Method}} & \emph{\textbf{Mean}} & \emph{\textbf{Standard deviation}} \\ \hline
 Proposed method & $0.6998$ & $0.3130$ \\ \hline
 Only DR & $0.9497$ & $0.3393$ \\ \hline
 Only MBC & $0.6456$ & $0.5002$ \\ \hline
 No control & $2.0325$ & $1.2091$ \\ \hline
\end{tabular}
    }
\label{Table_Result_MonteCarlo}
\end{center}
\end{table}

\begin{table}[htbp]
\caption{$2$-norm of the tracking error in the extreme-parameter evaluation (five seeds)}
\begin{center}
    {\renewcommand\arraystretch{1.2}
        \tabcolsep = 8.0pt
\begin{tabular}{|c|c|c|}
\hline
 \emph{\textbf{Method}} & \emph{\textbf{Mean}} & \emph{\textbf{Standard deviation}} \\ \hline
 Proposed method & $1.0911$ & $0.4104$ \\ \hline
 Only DR & $1.4778$ & $0.6627$ \\ \hline
 Only MBC & $1.6698$ & $1.3105$ \\ \hline
 No control & $3.5999$ & $1.7109$ \\ \hline
\end{tabular}
    }
\label{Table_Result_ExtremeParameter}
\end{center}
\end{table}

The most important finding from these results is that the proposed method exhibits the most consistent robustness, with the smallest variability, across diverse plant variations. This is considered to result from the synergy between the baseline performance maintained by the MBC, which eases exploration during training, and the generalization ability provided by DRL. Only MBC achieves the smallest mean norm in Table~\ref{Table_Result_MonteCarlo}, but its standard deviation is large, so that its performance varies with the plant realization. As Table~\ref{Table_Result_ExtremeParameter} shows, this is because its performance deteriorates severely, particularly near the extreme parameter combinations. Only DR exhibits smaller variability than Only MBC but insufficient performance, because training does not progress sufficiently as DR enlarges the region to be explored. The proposed method achieves the best performance in Table~\ref{Table_Result_ExtremeParameter}; that is, DRL imparts robustness even to the plants for which the MBC deteriorates severely.

\section{Experimental Verifications}   \label{S_Experiment}

\subsection{Experimental Setup and Conditions}   \label{S_Experiment_Setup}
\begin{figure}[htbp]
    \centering
    \includegraphics[width=0.38\textwidth]{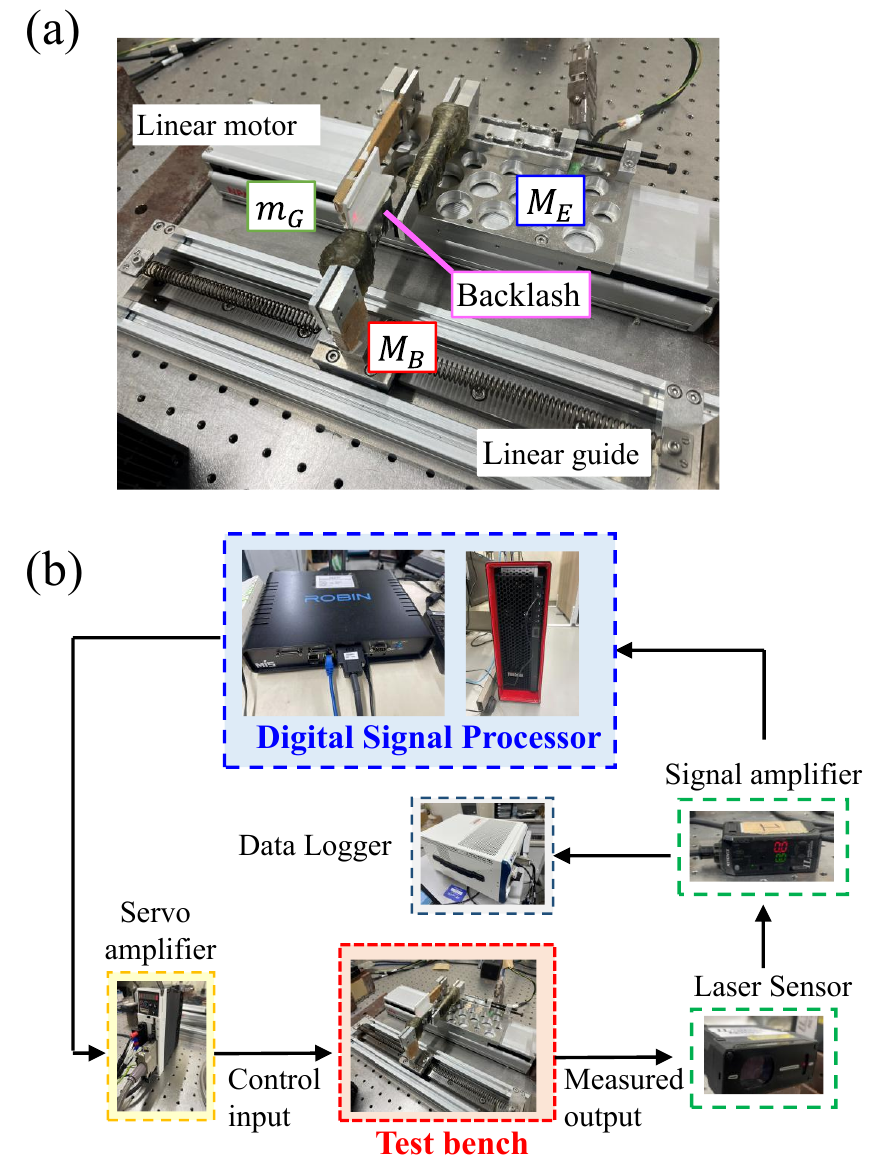}
    \caption{Experimental setup. (a) Powertrain test bench. (b) Configuration of the closed-loop experiments.}
    \label{Fig_Experiment}
\end{figure}

The test bench used in the experiments is shown in Fig.~\ref{Fig_Experiment}(a). It reproduces the essential dynamics of a vehicle powertrain with backlash and has been employed in our previous studies \cite{Yonezawa2019a,Yonezawa2019b,Yonezawa2024a}. The bench is a three-mass system with the structure shown in Fig.~\ref{Fig_Powertrain}, driven by a linear motor whose thrust serves as the control input. The backlash is realized as an adjustable clearance between the leaf springs on both sides of the intermediate mass. Further details of the bench are given in \cite{Yonezawa2019a,Yonezawa2024a}.

Fig.~\ref{Fig_Experiment}(b) illustrates the configuration of the closed-loop experiments. The four control systems compared in Section~\ref{S_Numerical_Verifications}, namely No control, Only MBC, Only DR, and the proposed method, are implemented on a DSP (MIS Corporation) with a control period of $\qty{6}{ms}$, which coincides with the sampling time in Table~\ref{Table_TD3_Parameter}. The policies, i.e., multilayer perceptrons with two hidden layers of 64 neurons each, are executed on the DSP together with the MBC. The policies trained in simulation are deployed as they are, without any additional training or tuning on the hardware.

The parameters used for training and for the design of the MBC are fixed at the nominal values used in Section~\ref{S_Numerical_Verifications}, and the discrepancies are introduced on the hardware side. Two cases are considered. In Case~1, the backlash length is set to the value of $\qty{5}{mm}$, and a reference profile representing abrupt changes in the driving force is applied. In Case~2, the backlash length is increased by 50\% to $\qty{7.5}{mm}$, which corresponds to the upper bound $\delta^{max}$ of the randomized range in Table~\ref{Table_Randomized_Parameters}, and a reference profile different from that of Case~1 is applied. In addition, the damping characteristics of the bench, which are difficult to quantify, are expected to deviate from the nominal values owing to aging and other factors; both cases are therefore subject to such deviations.

For the learning-based methods, all five policies trained with different seeds are tested on the hardware, and the $2$-norm of the tracking error is reported as the mean and standard deviation over the five policies.

\subsection{Experimental Results and Discussion}   \label{S_Experiment_Results}
The experiments examine whether the DRL policies trained in the simulation environment can be transferred to the physical system.

\begin{figure}[htbp]
    \centering
    \includegraphics[width=\columnwidth]{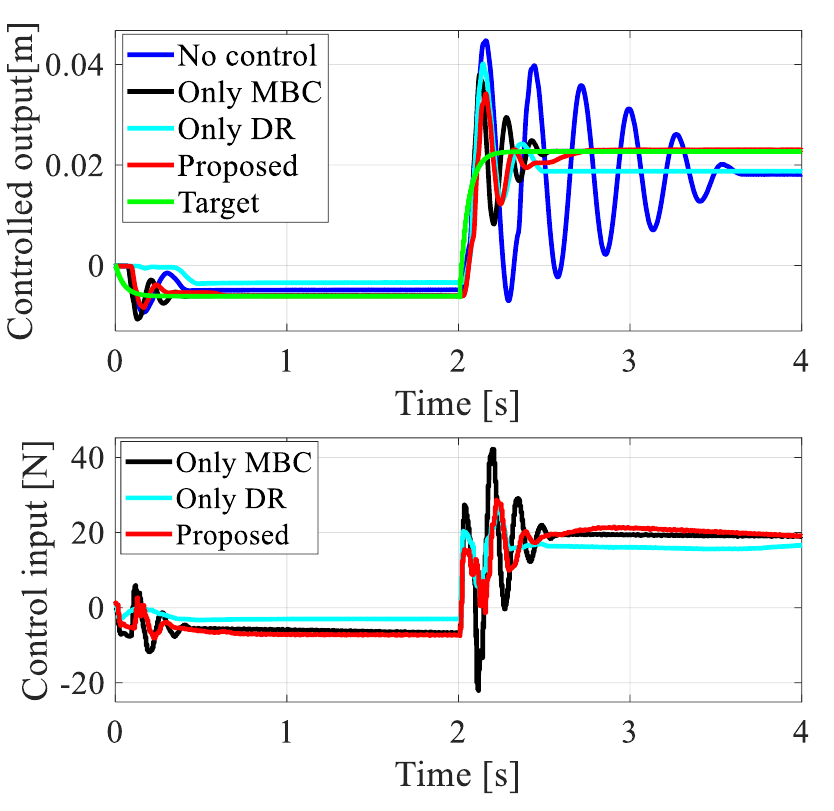}
    \caption{Experimental results in Case~1.}
    \label{Fig_Robustness_Case1}
\end{figure}

\begin{figure}[htbp]
    \centering
    \includegraphics[width=\columnwidth]{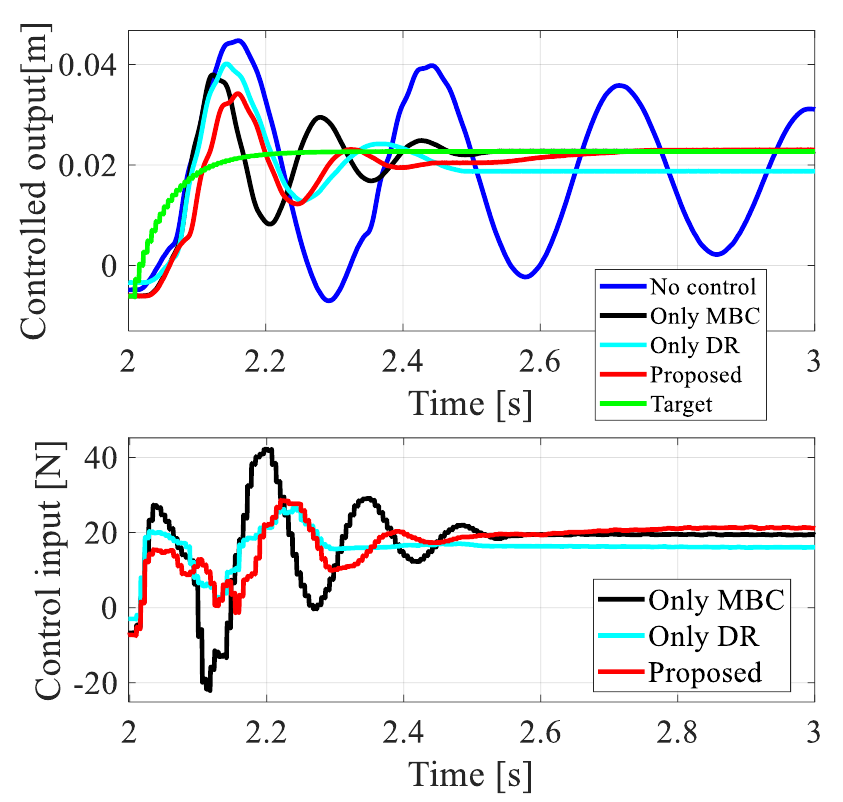}
    \caption{Enlarged view of the results in Case~1.}
    \label{Fig_Robustness_Case1_Enlarged}
\end{figure}

\begin{figure}[htbp]
    \centering
    \includegraphics[width=\columnwidth]{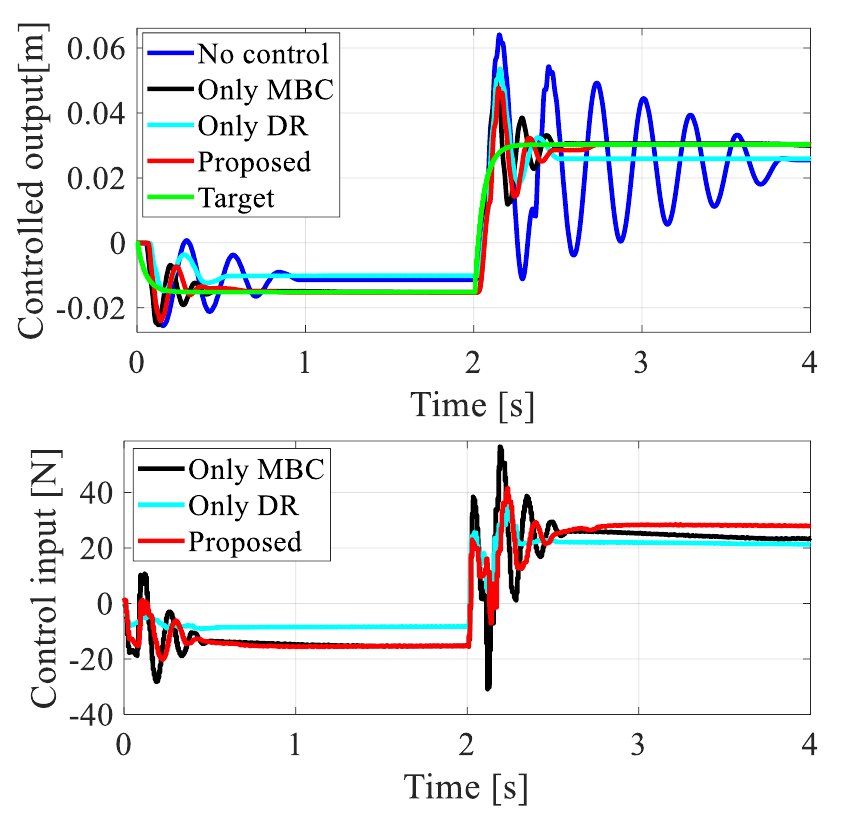}
    \caption{Experimental results in Case~2.}
    \label{Fig_Robustness_Case2}
\end{figure}

Figs.~\ref{Fig_Robustness_Case1} and~\ref{Fig_Robustness_Case1_Enlarged} show the experimental results in Case~1 and an enlarged view of them, respectively, and Fig.~\ref{Fig_Robustness_Case2} shows the results in Case~2. The $2$-norms of the tracking error in Cases~1 and~2 are summarized in Tables~\ref{Table_Result_Case1} and~\ref{Table_Result_Case2}, respectively.

As these results show, the proposed method (red line) achieves the most successful sim-to-real transfer, which is quantitatively supported by the smallest $2$-norms in Tables~\ref{Table_Result_Case1} and~\ref{Table_Result_Case2}. In Fig.~\ref{Fig_Robustness_Case1_Enlarged}, the proposed method suppresses the overshoot and the transient vibration immediately after the abrupt change in the driving force more effectively than the other closed-loop systems, and its response is the closest to the reference. The test bench is subject to the strong nonlinearity of backlash and, as noted in Section~\ref{S_Experiment_Setup}, to deviations of the damping characteristics due to aging; the results show that these uncertainties are compensated for to some extent. In other words, the generalization ability that DRL adds to the MBC is confirmed on the physical system. In contrast, Only MBC leaves larger transient vibration, which stems from uncertainties not accounted for in the controller design. In particular, the shock caused by the collision in the backlash affects its response more prominently.

\begin{table}[htbp]
\caption{$2$-norm of the tracking error in Case~1 (mean and standard deviation over five seeds)}
\begin{center}
    {\renewcommand\arraystretch{1.2}
        \tabcolsep = 8.0pt
\begin{tabular}{|c|c|c|}
\hline
 \emph{\textbf{Method}} & \emph{\textbf{Mean}} & \emph{\textbf{Standard deviation}} \\ \hline
 Proposed method & $0.2339$ & $0.0019$ \\ \hline
 Only DR & $0.4000$ & $0.0021$ \\ \hline
 Only MBC & $0.2631$ & -- \\ \hline
 No control & $0.7391$ & -- \\ \hline
\end{tabular}
    }
\label{Table_Result_Case1}
\end{center}
\end{table}

The response of Only DR (cyan line) is characterized by a conservative control input. Its large overshoot indicates that the compensation for backlash learned in simulation does not transfer well to the hardware. Although other uncertainties specific to the hardware, discussed later, also contribute to this result, factors arising from training cannot be ignored. When training relies on DRL alone, the exploration range enlarged by DR is considered too broad for the compensation of each individual uncertainty to be learned sufficiently. The proposed method, by contrast, secures the baseline performance from the outset and thereby promotes the acquisition of robustness focused on the uncertainties.

The control input highlights another benefit of introducing DRL, namely control efficiency. Compared with the proposed method and Only DR, Only MBC generates unnecessarily large control inputs. This inefficiency results from modeling errors, in other words, from a design that assumes a single fixed plant. In training with DR, by contrast, the agent learns how to adjust the control input while experiencing various plants, which is considered to lead to more efficient control. In particular, the DRL component of the proposed method corrects the excessive input of the MBC toward a more efficient one. The control input is analyzed in detail below.

\begin{table}[htbp]
\caption{$2$-norm of the tracking error in Case~2 (mean and standard deviation over five seeds)}
\begin{center}
    {\renewcommand\arraystretch{1.2}
        \tabcolsep = 8.0pt
\begin{tabular}{|c|c|c|}
\hline
 \emph{\textbf{Method}} & \emph{\textbf{Mean}} & \emph{\textbf{Standard deviation}} \\ \hline
 Proposed method & $0.3605$ & $0.0039$ \\ \hline
 Only DR & $0.5455$ & $0.0018$ \\ \hline
 Only MBC & $0.3685$ & -- \\ \hline
 No control & $1.1189$ & -- \\ \hline
\end{tabular}
    }
\label{Table_Result_Case2}
\end{center}
\end{table}

\begin{figure}[htbp]
    \centering
    \includegraphics[width=0.36\textwidth]{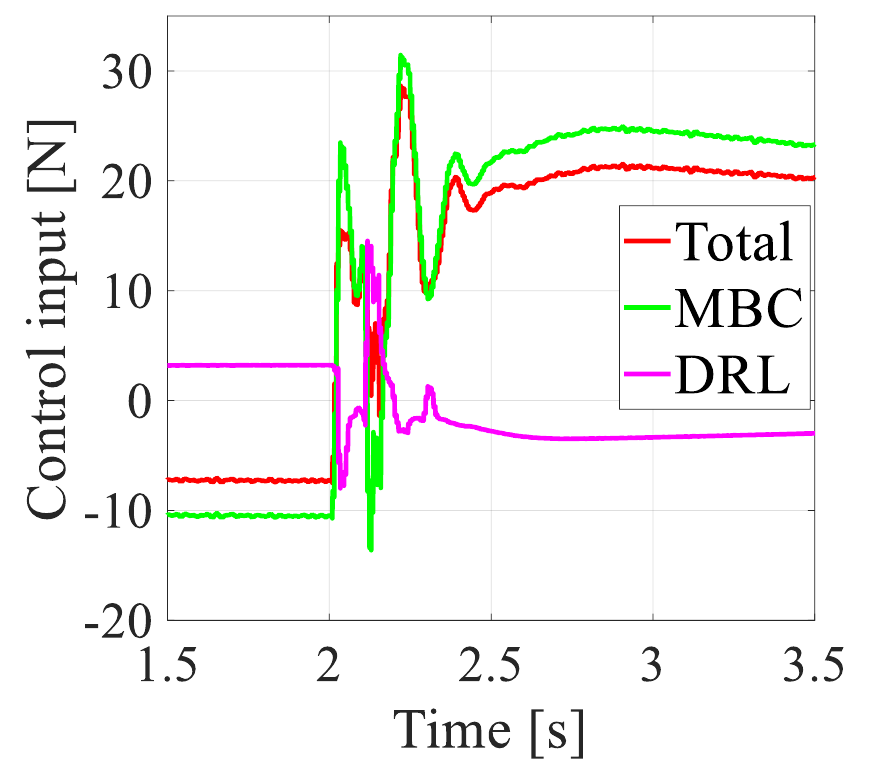}
    \caption{Decomposition of the control input of the proposed method into the contributions of the MBC and DRL.}
    \label{Fig_Input}
\end{figure}

To examine the composition of the control input, Fig.~\ref{Fig_Input} shows the contributions of the MBC and DRL separately. The MBC input (green line) is in phase with the total control input (red line), indicating that the MBC dominates the overall behavior. The DRL input (magenta line), in turn, corrects the excessive commands of the MBC and shapes them appropriately, that is, so as to compensate for the uncertainties. Immediately after $\qty{2.0}{s}$, a clearance momentarily opens in the backlash. If the excessive input of the MBC were applied as it is, a shock would occur at the subsequent collision. To prevent this, the collision should be cushioned, that is, a soft landing is desired. Around $\qty{2.0}{s}$, DRL computes an input in antiphase with that of the MBC, which decelerates the actuator and thereby compensates for the backlash. This complementary interaction between DRL and the MBC was identified through a numerical analysis in our previous work \cite{Yonezawa2025}; one of the novelties of this study is that its existence is confirmed experimentally.

The experimental results also reveal room for improvement in control performance. In particular, the proposed method does not suppress the overshoot as effectively as in the simulation; compared with the ideal response (green line), the compensation for backlash is still insufficient. Moreover, unlike the trend in the simulation, the standard deviation of the proposed method in Table~\ref{Table_Result_Case2} is larger than that of Only DR. These observations are attributed to uncertainties in the hardware that are not considered in the training environment; specifically, additional friction in the actuator and nonlinear characteristics of the springs may be present. The steady-state error of Only DR suggests a dominant influence of friction or similar effects. With respect to the steady-state error, introducing an MBC that explicitly includes integral action is effective, as demonstrated by the proposed method. On the other hand, the overshoot and residual vibration caused by backlash may be attributed to the degradation of the MBC performance. Accordingly, one limitation of this study is its vulnerability to uncertainties not assumed in the training environment. Possible countermeasures include improving the robustness of the MBC and refining the training environment to reduce the sim-to-real gap.

\section{Conclusion}   \label{S_Conclusion}
This paper presented MBCADR, a framework that integrates model-based control into domain-randomized DRL for robust vibration suppression of powertrains with backlash. A residual policy trained by TD3 under randomized dynamics compensates for the model mismatch of an $H_{2}$ controller designed from the nominal model. Domain-randomized training was formulated as policy optimization over a family of MDPs, and a Lyapunov-based analysis provided a forward-invariant set and an ultimate bound for the hybrid closed loop. Monte Carlo and extreme-parameter evaluations over five training seeds showed that the proposed method yields the smallest performance variability across plant realizations. Experiments on a powertrain test bench confirmed that the policies trained in simulation transfer to the hardware without further tuning. Future work includes applying the framework to other control systems and further mitigating the conservativeness induced by domain randomization.

\section*{Acknowledgment}
During the preparation of this work, the author(s) used Claude Opus 5.5 (Anthropic) for language editing, drafting, and assistance with manuscript preparation. After using this tool/service, the author(s) reviewed, verified, and edited the content as needed and take(s) full responsibility for the content of the published article.

\bibliographystyle{IEEEtran}   
\bibliography{References_CCTA2026}  

\vfill

\end{document}